\documentclass[12pt,preprint]{aastex}
\newcommand{\cdi}{\ensuremath{\cos\Delta_i}}
\newcommand{\sdi}{\ensuremath{\sin\Delta_i}}
\newcommand{\partials}[2]{\ensuremath{\partial Q^2\over\partial #1_#2}}
\shortauthors{JENKINS}
\shorttitle{THERMAL PRESSURES IN THE LOCAL BUBBLE}
\received{2002 May 21}

\begin{document}

\title{Thermal Pressures in Neutral Clouds inside the Local Bubble, as
Determined from C~I Fine-Structure Excitations\altaffilmark{1}}

\altaffiltext{1}{Based on
observations with the NASA/ESA {\it Hubble Space Telescope\/} obtained
at the Space Telescope Science Institute, which is operated by the
Association of Universities for Research in Astronomy, Inc., under NASA
contract NAS 5-26555.}

\author{Edward B. Jenkins}

\affil{Princeton University Observatory\\
Princeton, NJ 08544-1001}

\email{ebj@astro.princeton.edu}
\begin{abstract}
High-resolution spectra covering the absorption features from
interstellar C~I were recorded for four early-type stars with
spectrographs on the {\it Hubble Space Telescope}, in a program to
measure the fine-structure excitation of this atom within neutral clouds
inside or near the edge of the Local Bubble, a volume of hot ($T\sim
10^6\,$K) gas that emits soft x-rays and extends out to about 100~pc
away from the Sun.  The excited levels of C~I are populated by
collisions, and the ratio of excited atoms to those in the ground level
give a measure of the local thermal pressure.  Absorptions from the two
lowest levels of C~I were detected toward $\alpha$~Del and $\delta$~Cyg,
while only marginal indications of excited C~I were obtained for
$\gamma$~Ori, and $\lambda$~Lup.  Along with temperature limits derived
by other means, the C~I fine-structure populations indicate that for the
clouds in front of $\gamma$~Ori, $\delta$~Cyg and $\alpha$~Del,
$10^3<p/k<10^4\,{\rm cm}^{-3}\,$K at about the $\pm 1\sigma$ confidence
level in each case.  The results for $\lambda$~Lup are not as well
constrained, but still consistent with the other three stars.  The
results indicate that the thermal pressures are below generally accepted
estimates $p/k>10^4\,{\rm cm}^{-3}\,$K for the Local Bubble, based on
the strength of x-ray and EUV emission from the hot gas.  This
inequality of pressure for these neutral clouds and their surroundings
duplicates a condition that exists for the local, partly-ionized cloud
that surrounds the Sun.  An appendix in the paper describes a direct
method for determining and eliminating small spectral artifacts arising
from variations of detector sensitivity with position.
\end{abstract}

\keywords{ISM: atoms -- ISM: lines and bands -- techniques:
spectroscopic -- ultraviolet: ISM}

\section{Introduction}\label{intro}

The Sun is immersed in two concentric volumes of interstellar material
with very different properties.  In the immediate surroundings [out to
distances ranging from 0.05 to several pc away  (Redfield \& Linsky
2000)], the gas is partly ionized, with characteristic properties $0.10
< n({\rm H~I}) < 0.24\,{\rm cm}^{-3}$, $0.04 < n(e) < 0.10\,{\rm
cm}^{-3}$ and $T\approx 7500\,$K  (Lallement 1998; Puyoo \& Ben Jaffel
1998; Ferlet 1999; Gry \& Jenkins 2001).  This cloud, often called the
Local Interstellar Cloud (LIC), is surrounded by a large cavity
containing fully-ionized, very hot, low-density material with $n({\rm
H}^+)\approx 0.008\,{\rm cm}^{-3}$ and $T\sim 10^6\,$K  (Bergh\"ofer et
al. 1998; Burrows \& Guo 1998) that is prominent in soft x-ray emission 
(Snowden et al. 1990, 1998).  This large volume of hot gas is believed
to be the product of probably several supernova explosions  (Cox \&
Reynolds 1987; Breitschwerdt \& Schmutzler 1994; Ma\'iz-Apell\'aniz
2001; Smith \& Cox 2001) and is known as the Local Bubble (LB).  The LB
is also conspicuous by showing a general absence of cold, neutral gas up
to a well defined boundary  (Sfeir et al. 1999; Vergely et al. 2001), as
is shown in Figure~\ref{sfeirplt}.

A persistent puzzle has been the apparent mismatch in thermal pressures
between the two media.  Using the parameters stated above, one finds
that $1400 < p/k < 3600\,{\rm cm}^{-3}\,$K for the LIC, which contrasts
with an apparent representative value of $16,000\,{\rm cm}^{-3}\,$K for
the fully-ionized, hot gas in the LB.  The objective of the research
presented here is to help answer the question, ``What thermal pressures
are found for other neutral clouds within the Local Bubble?  That is, do
they have values similar to the LIC, or are they closer to matching the
apparent pressure of the LB?''  To gain insights on this question, one
can observe stars located behind individual clouds inside the LB and
measure their absorption features of C~I arising from different
fine-structure levels of the ground state.  The ratios of populations of
these states are governed by an equilibrium between collisional
excitations (governed by local densities and temperatures) and
spontaneous radiative decays.

Stars suitable for viewing the C~I features had to satisfy a number of
conditions to yield useful results.  Their selection is described in
\S\ref{plan}; ultimately four such stars were observed in a manner
described in \S\ref{obs}.  The C~I absorption features are very weak,
and in order to obtain useful measures of their strengths particular
care was exercised to remove spurious signals arising from the detector,
as outlined in \S\ref{artifacts}.  (Mathematical details about the
correction method are presented separately in the Appendix.) 
Section~\ref{analysis} describes how the equivalent widths of various
absorption features were combined and corrected for saturation (very
mild, except for one of the stars).  This section also discusses the
derivations of fine-structure population ratios, which may be compared
to the theoretically expected values for different conditions
(\S\ref{expected_f1}).  Before one can derive useful limits for the
thermal pressures, the allowable ranges of temperature must be
constrained, and different methods of deriving these constraints are
discussed in \S\ref{temp_limits}.  Ultimately, the limits for the
population ratios and temperatures restrict the possible values for
$p/k$ [and local density $n({\rm H})$], as shown for the four cases in
the diagrams presented in Fig.~\ref{limit_panels}.  Three out of the
four stars indicate internal thermal pressures for the foreground clouds
that are below the generally accepted value for the LB.  Possible
explanations for this imbalance, duplicating that seen for LIC, are
presented in \S\ref{summary}.

\section{Selection of Target Stars}\label{plan}

Target stars chosen for the survey had to satisfy four fundamental
criteria.  First, the stars had to be within the Local Bubble or near
its edge.  Second, they had to have sufficient neutral gas in front so
that there was a reasonable expectation of seeing the C~I absorption
features.  Third, the stars had to be bright enough to give a good
signal-to-noise ratio in a reasonably short observing time.  Finally,
the survey avoided stars with projected rotational velocities $v\sin i <
50\,{\rm km~s}^{-1}$  (Uesugi \& Fukuda 1981 -- with actual values
provided by the VizieR web site at the Strasbourg Data Center), so that
stellar features would not cause confusion when the interstellar lines
were being identified and measured.  To satisfy the first two
requirements, the selection included stars less than about 100~pc away
that had interstellar D-line absorption features indicating
$10^{11}<N$(Na~I) $< 3\times 10^{11}\,{\rm cm}^{-2}$.  A  compilation by
Welsh et al  (1994) was a good source of information about D-line
absorption at the time the survey was being planned.  Fluxes at
1565\,\AA\ listed in the {\it TD-1 Catalogue of Stellar Ultraviolet
Fluxes\/}  (Thompson et al. 1978) provided a good guide for selecting
targets with satisfactory brightness levels.

To increase the chance that the C~I features arose from truly
intervening material rather than circumstellar disks or shells around
the target stars, the survey did not include candidates that had
spectral types with an emission-line ``e'' designation  (Hoffleit \&
Jaschek 1982; Slettebak 1982) or an excess IRAS flux at $12\mu$m
relative to the normal expectation.  Short-period binaries were also
rejected, since they could have interacting gas streams.  As a final
precaution against skewing the results with cases dominated by gaseous
material very near the stars, there was an exclusion of targets that had
excess diffuse infrared emission nearby in the sky  (Gaustad \& Van
Buren 1993), signifying the possible presence of dust grains being
heated by the star.

\begin{figure}
\plotone{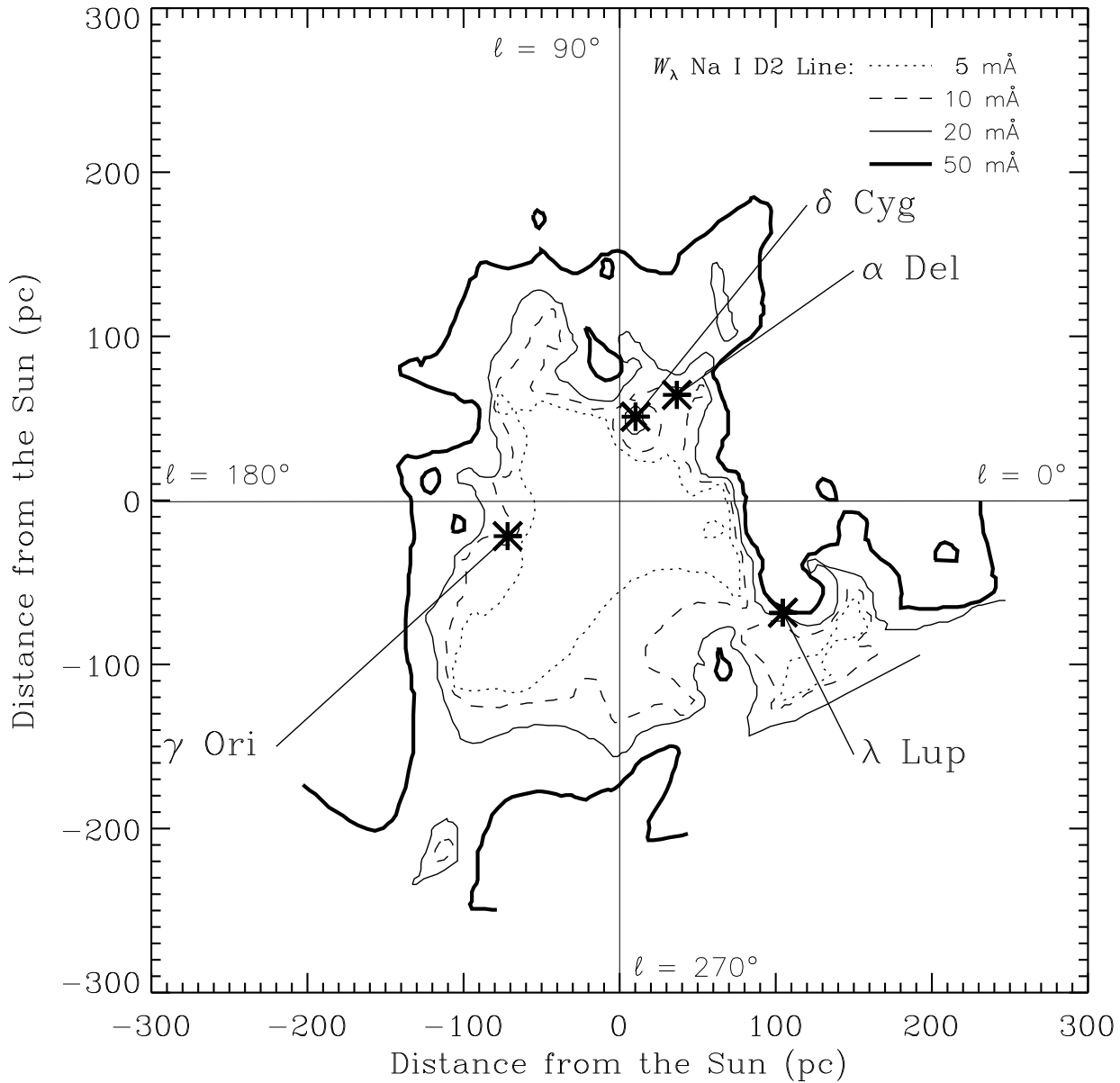}
\caption{Locations of the four program stars listed in
Table~\protect\ref{target_stars} with respect to the shape of the Local Bubble in
the plane of the Galaxy, outlined by contours of increasing Na~I D2-line
absorption equivalent widths mapped by Sfeir et al  (1999).  The Galactic center
direction is toward the right.  To maintain consistency with Sfeir et al's
construction of the contours according to a spherical projection, the distances of
the stars from the origin are not foreshortened by $\cos (b)$.\label{sfeirplt}}
\end{figure}

Four stars that were ultimately chosen for the survey are listed in
Table~\ref{target_stars}.  Their locations with respect to the Local
Bubble boundaries mapped by Sfeir et al.  (1999) are shown in
Figure~\ref{sfeirplt}.

\placefigure{sfeirplt}
\placetable{target_stars}
\begin{deluxetable}{
r    
r    
r    
r    
r    
c    
l    
}
\tablecolumns{7}
\tablewidth{0pt}
\tablecaption{Target Stars and Dataset Identifications\label{target_stars}}
\tablehead{\colhead{Star} & \colhead{HD} & \colhead{$\ell$} &
\colhead{$b$} & \colhead{$d$\tablenotemark{a}} & \colhead{Spectral} & \colhead{HST
Archive}\\
\colhead{Name} & \colhead{Nr.} & \colhead{(deg)}& \colhead{(deg)} &
\colhead{(pc)} & \colhead{Type\tablenotemark{b}} & \colhead{Dataset Name}}
\startdata
$\gamma$~Ori&35468&196.9&$-16.0$&$75\pm 6$&B2$\,$III&Z3CI0107T\\
$\lambda$~Lup&133955&326.8&+11.1&$125\pm 11$&B3$\,$V&Z3CI0307T\\
$\delta$~Cyg&186882&78.7&+10.2&$52\pm 1$&B9.5$\,$IV&Z3CI0407T\\
$\alpha$~Del&196867&60.3&$-15.3$&$74\pm 4$&B9$\,$IV&O50001010-40\\
\enddata
\tablenotetext{a}{From the parallax measurements of Hipparcos  (Perryman et al.
1997) reported on the VizieR web site at the Strasbourg Data Center.}
\tablenotetext{b}{Hoffleit \& Jaschek  (1982).}
\end{deluxetable}
\clearpage

\section{Observations}\label{obs}

The Goddard High Resolution Spectrograph (GHRS) on HST had a proven
capability of achieving large signal-to-noise ratios for bright targets 
(Cardelli \& Ebbets 1994; Cardelli 1995).  For this reason, this
instrument was considered ideal for the objective of recording the weak
C~I absorption features.  The survey was originally intended to be
completed before GHRS was to be replaced by the Space Telescope Imaging
Spectrograph (STIS) during the second HST servicing mission, but useful
observations of $\alpha$~Del were not accomplished until after STIS was
installed. 

Observations of $\gamma$~Ori, $\lambda$~Lup and $\delta$~Cyg were
carried out with the Ech-A grating on GHRS, with a substep pattern that
created 4 spectrum bins per detector diode (STEP-PATT=7) plus a sampling
of the background level.  The standard option for spectrum shifts
(FP-SPLIT mode) allowed for an elimination of the detector's
fixed-pattern noise in the spectrum (\S\ref{artifacts} and
Appendix~\ref{fpn_analysis}).  The GHRS observations were performed with
COSTAR in place to correct for the telescope's spherical aberration, and
the 2\arcsec\ entrance aperture was used to increase the efficiency. 
The STIS observations of $\alpha$~Del with the E140H grating had to be
done using neutral density filters to prevent the MAMA detector count
rate limits from being exceeded.  The wavelength resolving powers of the
GHRS observations were $\lambda/\Delta\lambda = 90,000$ [but with some
extended wings in the line-spread function -- see Howk, Savage \& Fabian 
(1999)], while that for the STIS spectrum of $\alpha$~Del is 110,000 
(Kimble et al. 1998; Leitherer 2001).

The limited wavelength coverage of the GHRS Digicon detector for echelle
spectroscopy limited observations to only one multiplet at any
particular moment.  The multiplet of choice was the one at 1260$\,$\AA,
since it had well separated lines and was near the peak in the
instrument's sensitivity.  However, for late B-type stars ($\delta$~Cyg
and $\alpha$~Del), a strong stellar feature seriously depresses the flux
level at the position of this multiplet, so other multiplets had to be
viewed for these stars.  The STIS observation of $\alpha$~Del covered
many multiplets, but only the strong ones listed in Table~\ref{EW} gave
useful results.  For all stars, the S~II triplet with features at
1250.584, 1253.811 and 1259.519$\,$\AA\ was also observed to monitor the
presence of a generally undepleted element in its favored stage of
ionization in an H~I region.  The abundance of S~II served a useful
purpose in constraining the temperature of the gas, as outlined in
\S\ref{temp_limits}. 

\section{Data Reduction}\label{data_reduction}

\subsection{Removal of Detector Artifacts}\label{artifacts}

In order to realize the full potential of GHRS in sensing very weak
absorption features, it was necessary to remove spurious signal
deviations caused by small changes in photocathode sensitivity with
position.  The intentional displacements in wavelength for different
subexposures in the GHRS FP-SPLIT observing routine allowed these
variations to be differentiated from real absorption features in the raw
spectra.  The method used here to derive independently the detector's
fixed-pattern signal and the real spectral signal gave a direct solution
and thus differed from the iterative technique described by Ebbets 
(1992), Cardelli \& Ebbets  (1994) and Fitzpatrick \& Spitzer  (1994). 
Details of this more direct method of solution are given in
Appendix~\ref{fpn_analysis}.

\begin{figure}
\plotone{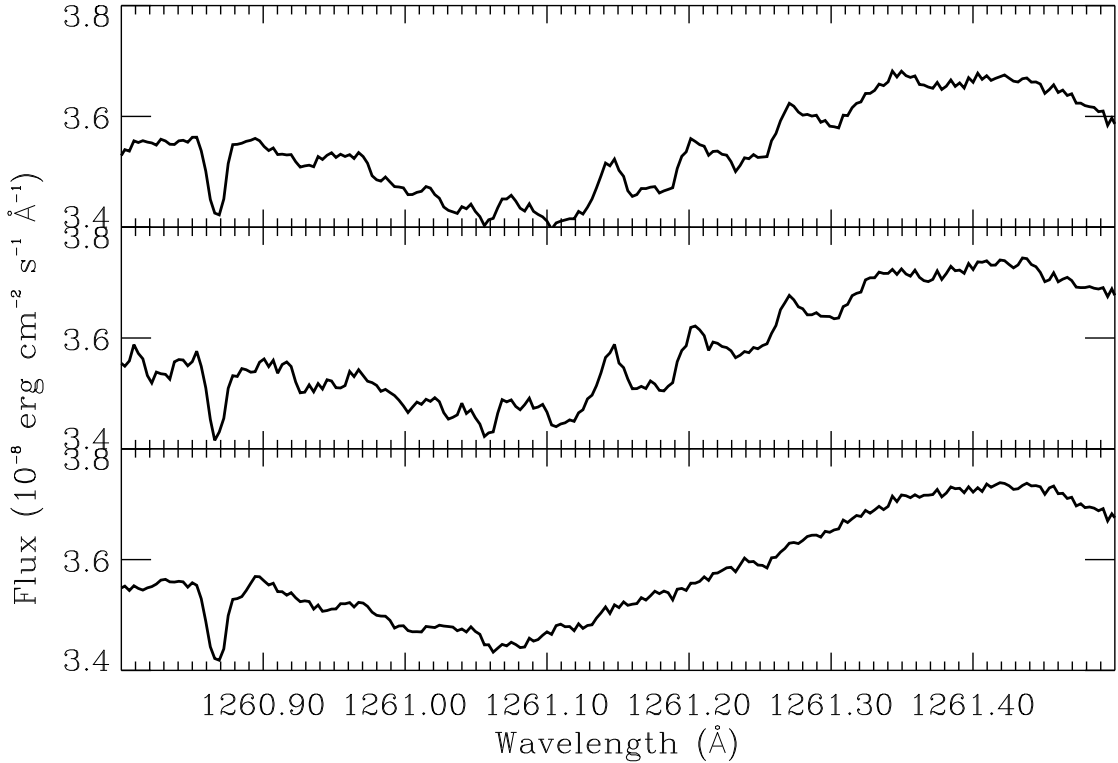}
\caption{Different reductions applied to the spectrum of $\gamma$~Ori: (top panel)
a simple co-addition of the individual spectra with shifts that line up the
spectral features, (middle panel) a composite spectrum corrected for pattern noise
assuming that it is fixed with respect to the detector diodes, using the method
described in Appendix~\protect\ref{fpn_analysis}, and (bottom panel) once again a
spectrum corrected for the detector pattern, but this time with offsets of up to
$\Delta\lambda=0.0348$\AA\ ($8.3\,{\rm km~s}^{-1}$) to compensate for the fact
that the strongest flaws move with respect to the diodes from one exposure to the
next.  See \S\protect\ref{artifacts} for details.  The absorption feature on the
left-hand side is the 1260.736$\,$\AA\ transition of C~I at a heliocentric
velocity of $+31\,{\rm km~s}^{-1}$ (uncalibrated for small wavelength errors in
the spectrograph).  A strong transition of C~I* at 1261.122$\,$\AA\ is too weak to
be clearly seen (it should appear at 1261.252$\,$\AA\ on the wavelength scale
presented here).\label{demofig}}
\end{figure}

It is important to recognize that for any given wavelength offset within
the FP-SPLIT procedure, there are two effects that can alter the mapping
of photocathode points onto given spectral bins.  One of them is a
change caused by the Doppler-shift correction for orbital motion
(performed internally by GHRS).  The extremes of this correction can in
principle differ by as much as twice the orbital velocity of
approximately $7\,{\rm km~s}^{-1}$ times the cosine of the angle of the
target with respect HST orbital plane at the time of observation.  The
other effect, a much smaller one, arises from small changes in the
influence of the Earth's magnetic field on the trajectories of the
electrons traveling from the detector's photocathode to the sensing
diodes  (Ebbets 1992).  Figure~\ref{demofig} illustrates the importance
of recognizing these shifts and compensating for them in the analysis
that eliminates the fixed-pattern noise.  The top panel shows a straight
average of the spectra of $\gamma$~Ori (i.e., no attempt to correct for
the fixed-pattern noise), after shifts had been implemented to
compensate for spectral motions with respect to the diodes.  The middle
panel shows the outcome for an initial attempt to correct for the
fixed-pattern variations by assuming that they remained stationary with
respect to the pixel assignments in the accumulated signal transmitted
to the ground.  Practically no improvement over the simple average is
apparent after this correction.  However, one may track the movement of
the photocathode by measuring the shifts of one or more particularly
strong flaws or, if they are not apparent, by cross correlating the
spectra.  After this is done and the spectra have had their
fixed-pattern features aligned, an analysis (with appropriate
modifications in the spectral shifts) produces a satisfactory result
(bottom panel).  However, at the point that we declare the pattern to be
fixed with respect to the photocathode, we abandon our ability to
compensate for differences in response of the detector's diode elements. 
(By an extension of the analysis method presented in
Appendix~\ref{fpn_analysis}, it is possible in principle to correct for
both sources of variation.)

\placefigure{demofig}

The fixed pattern removal is beneficial only when the fluctuations are
larger than the random noise arising from the counting of photoevents. 
For each spectrum a check was made to insure that this condition
applied, and, if not, a simple coaddition was performed instead.  There
was only one instance (the coverage of the 1560$\,$\AA\ multiplet of C~I
for $\delta$~Cyg) where fluctuations in the fixed-pattern corrected
spectrum exceeded those in the simple coadded spectrum, and then by only
a small amount.

\subsection{Scattered Light Corrections}\label{scat_cor}

Echelle spectrographs create complex background light patterns arising
from scattering from the echelle and cross-disperser gratings 
(Cardelli, Ebbets, \& Savage 1993).  Scattered light corrections
supplied by the standard calibrated data products supplied by the Space
Telescope Science Institute were utilized for the GHRS spectra.  For the
STIS spectrum of $\alpha$~Del, the characterization of scattered light
developed by Lindler \& Bowers  (2000)  was invoked by utilizing the
CALSTIS reduction procedures developed for the STIS Investigation Team. 
Since the C~I features toward all of the stars in this survey are weak,
errors in the scattered light corrections have a negligible influence on
the results.

\subsection{Continuum Definitions}\label{cont_def}

Least-squares fits of Legendre polynomials to intensities on either side
of the absorption features defined the reference continuum levels for
the line measurements.  The procedures of Sembach \& Savage  (1992) were
adopted for determining the appropriate order of the polynomial. 
However, from mock continuum fitting exercises where no interstellar
features were present, it appeared that terms in the error matrix for
the polynomial coefficients generally underestimated the true
uncertainties in the outcomes by about a factor of two.  The additional
deviations probably arise from errors in the assumptions that
polynomials are truly appropriate for describing stellar continuum
levels over arbitrary wavelength intervals.  Thus, to account for the
uncertainties that are likely to reach beyond the formal errors, the
$\pm 2\sigma$ deviations were declared to be really $\pm 1\sigma$
uncertainties.

\section{Analysis of the Absorption Features}\label{analysis}

\subsection{Column Densities}\label{col_densities}

\begin{figure}
\plotone{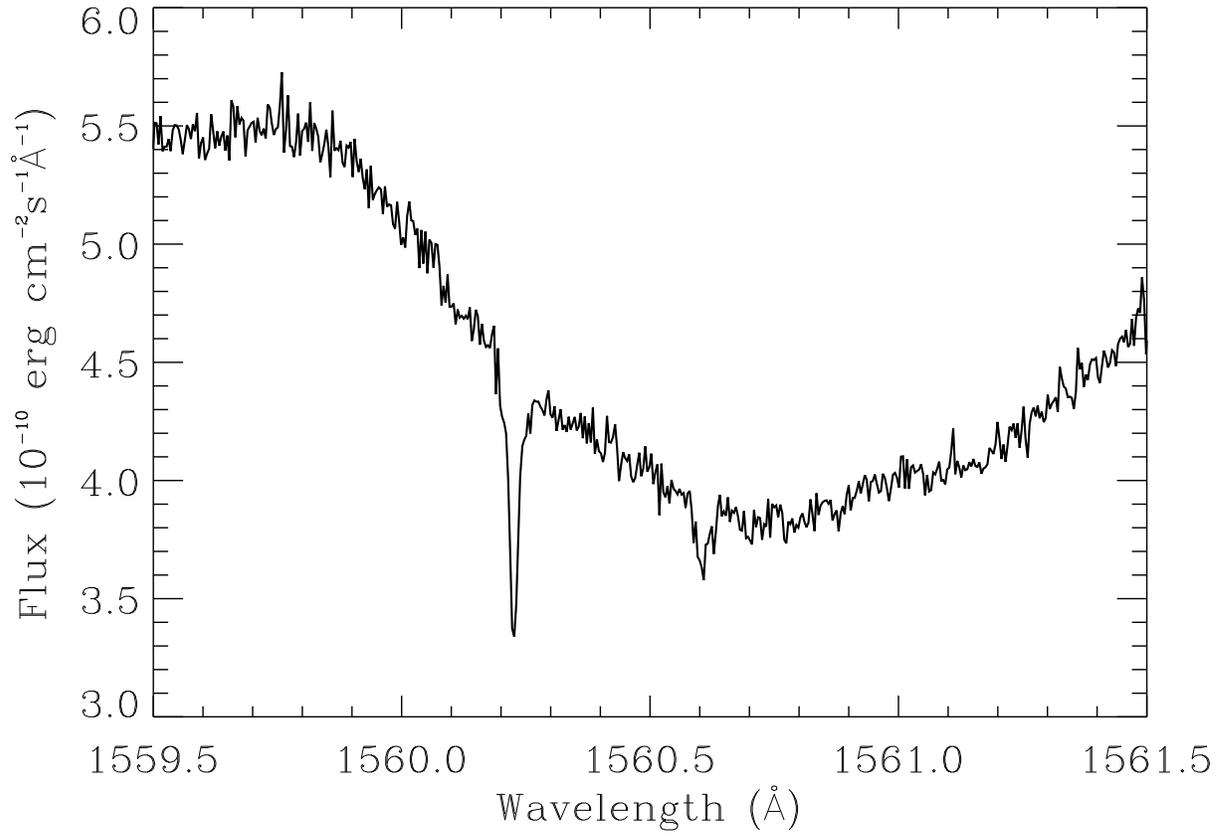}
\caption{Spectrum of $\delta$~Cyg covering the 1560$\,$\AA\ multiplet of C~I.  The
strong absorption feature is from the 1560.309$\,$\AA\ transition of C~I, while
the weaker one arises from a blend of two features from C~I* at 1560.682 and
1560.709$\,$\AA.\label{del_cyg}}
\end{figure}

Table~\ref{EW} lists the lines and their measured equivalent widths for
the carbon absorption features\footnote{Throughout this paper, the
designation ``C~I'' applies to neutral carbon atoms in the ground
$^3$P$_0$ state, whereas ``C~I*'' designates the first excited
fine-structure level ($^3$P$_1$).  Absorptions from the second excited
level, C~I**$\equiv^3$P$_2$, were too weak to observe.  The expression
${\rm C~I_{\rm total}}$ refers to neutral carbon in all stages of
excitation.} in the spectra of each of the four target stars.  An
example of a case ($\delta$~Cyg) where lines from both C~I and C~I*
could be seen is shown in Fig.~\ref{del_cyg}.

\placetable{EW}
\begin{deluxetable}{
r    
l    
l    
c    
c    
}
\tablecolumns{5}
\tablewidth{525pt}
\tablecaption{Absorption Lines and Equivalent Widths\label{EW}}
\tablehead{
\colhead{Star} & \colhead{Abs.} &
\colhead{$\lambda$} & \colhead{$\log f\lambda$\tablenotemark{a}} &
\colhead{W$_{\lambda}$}\\
\colhead{Name} & \colhead{State} & \colhead{(\AA)} & \colhead{ } &
\colhead{(m\AA)}
}
\startdata
$\gamma$~Ori&C~I&1260.736&1.870&\phs $0.85\pm 0.09$\\
&C~I*&1261.122\tablenotemark{b}&1.537&\phs $0.04\pm 0.07$\\
$\lambda$~Lup&C~I&1260.736&1.870&\phs $0.89\pm 0.13$\\
&C~I*&1260.927&1.517&\phs $0.24\pm 0.11$\\
&&1260.996&1.444&\phs $0.09\pm 0.10$\\
&&1261.122&1.537&$-0.03\pm 0.10$\\
$\delta$~Cyg&C~I&1560.309&2.099&\phs $6.25\pm 0.50$\\
&C~I*&1560.682; .709\tablenotemark{c}&2.099&\phs $1.92\pm 0.52$\\
$\alpha$~Del&C~I&1656.928&2.367&\phs $7.42\pm 0.86$\\
&&1560.309&2.099&\phs $3.80\pm 0.43$\\
&&1328.833&2.077&\phs $3.55\pm 0.75$\\
&&1277.245\tablenotemark{d}&2.225&\phs $2.68\pm 0.43$\\
&&&&\phs $3.26\pm 0.38$\\
&C~I*&1656.267&1.987&\phs $1.45\pm 0.84$\\
&&1657.379&1.765&\phs $0.43\pm 0.84$\\
&&1657.907&1.890&\phs $1.82\pm 0.73$\\
&&1560.682; .709\tablenotemark{c}&2.099&\phs $0.73\pm 0.74$\\
&&1329.085; .100; .123\tablenotemark{e}&2.166&\phs $0.30\pm 0.80$\\
&&1277.282\tablenotemark{d}&2.017&\phs $1.11\pm 0.45$\\
&&&&\phs $0.12\pm 0.42$\\
&&1277.513\tablenotemark{d}&1.703&\phs $0.52\pm 0.56$\\
&&&&$-0.14\pm 0.50$\\
\enddata
\tablenotetext{a}{Sources of $f$-values: multiplets at 1260 and 1329\AA\  from
Jenkins \& Tripp  (2001), multiplets at 1560 and 1657\AA\ from Wiese et al. 
(1996).}
\tablenotetext{b}{Other absorptions from C~I* in this multiplet were not measured
because they appeared in a small wavelength interval where fixed-pattern
corrections were not satisfactory enough to reveal weak absorption features.  See
Fig.\protect\ref{demofig}.}
\tablenotetext{c}{Two lines that are unresolved.}
\tablenotetext{d}{There are two opportunities to measure this feature, since it is
visible at opposite ends of two adjacent orders of diffraction in the STIS echelle
format.}
\tablenotetext{e}{Three lines that are unresolved.  The allowance for line
saturation was for a feature one-third the strength of this ensemble, since the
individual features are well separated compared to the real line widths and have
approximately the same strength.}
\end{deluxetable}
\placefigure{del_cyg}

Errors in equivalent widths arising from the continuum uncertainties
discussed above (\S\ref{cont_def}) were evaluated by comparing the
measurement outcomes for the most probable continuum levels with those
using reasonable departures for the continua (at the declared $\pm
1\sigma$ limits).  The resulting uncertainties in $W_\lambda$ should be
uncorrelated with errors arising from random noise in the signal inside
the measurement interval for the line.  Hence, in each case the
estimates for the amplitudes for these two different kinds of errors
could be added in quadrature.  The combined error predictions are
reflected in the uncertainties in equivalent widths listed in the last
column of Table~\ref{EW}.  In many cases, the measured values for
absorption out of C~I* were comparable to or weaker than their
associated errors, but these numbers were retained in the subsequent
analysis up to the point where the level populations are expressed (in
terms of a parameter called $f1$, as described later in this section),
in order to preserve the integrity of the final error estimates. 

While the lines are generally very weak, it is a mistake to assume that
under all circumstances they are either completely unsaturated or fully
resolved by the instrument.  Very high resolution spectra of Na~I
features observed with ground-based telescopes reveal features with
velocity dispersions that are extremely narrow.  Since both C~I and Na~I
represent ionization stages below those favored for H~I regions, their
abundances are driven in a very similar fashion by local conditions
(i.e., primarily the strength of ionizing radiation and local electron
density).  It is thus appropriate to use Na~I as a surrogate for C~I
when we wish to understand how the lines in the present study might
saturate.  The only complication in this comparison arises from the mass
difference of the two elements.  If turbulence is the dominant source of
broadening, the profiles of C~I and Na~I should be virtually identical. 
Conversely, if thermal broadening dominates, the C~I lines are broader
by a factor $(23/12)^{0.5}$.  In the analysis that produced column
densities of C~I, both extremes were considered.

Various investigators have determined the nature of the Na~I absorptions
toward the four stars in this study of C~I.  Vallerga et al  (1993)
observed a single absorption component with a velocity dispersion
$b=0.33\,{\rm km~s}^{-1}$ toward $\gamma$~Ori.  Crawford  (1991) found a
best fit to a single component in the spectrum of $\lambda$~Lup with
$b=1.5\,{\rm km~s}^{-1}$.  Both Blades, Wynne-Jones \& Wayte  (1980) and
Welty, Hobbs \& Kulkarni  (1994) report very similar values for $b$ for
a single component in the spectrum of $\delta$~Cyg (the former lists
$b=0.44\,{\rm km~s}^{-1}$ while the latter gives $b=0.42\,{\rm
km~s}^{-1}$).  The situation for $\alpha$~Del is more complex than the
others: Vallerga et al  (1993) found one component with $N$(Na~I) =
$9.5\times 10^{10}\,{\rm cm}^{-2}$ and $b=1.10\,{\rm km~s}^{-1}$ and
another, slightly overlapping feature with $N$(Na~I) = $1.6\times
10^{10}\,{\rm cm}^{-2}$ and $b=0.90\,{\rm km~s}^{-1}$.

Table~\ref{col_dens} lists the column densities of C~I and C~I* based on
curves of growth for the two extreme assumptions for the character of
the line broadening: (1) the line shapes are exclusively determined by
thermal Doppler motions at a temperature
\begin{equation}\label{T_NaI}
T={23m_pb_{\rm Na~I}^2\over 2k}
\end{equation}
so that $b_{\rm C~I}=(23/12)^{0.5}b_{\rm Na~I}$, and (2) the
temperatures are arbitrarily small and the broadening arises principally
from some form of bulk motion (e.g., turbulence), making the $b$-values
for the two elements identical.  For either of the two cases, the
curve-of-growth corrections were explicitly calculated for each line
using the nominal $W_\lambda$ values and their accompanying $\pm
1\sigma$ error limits.  When more than one transition could be used (or
a single transition was viewed more than once -- see Table~\ref{EW}),
the results reflect a weighted average, with weights proportional to
$\sigma_i^{-2}$ and a final error equal to $(\sum_i
\sigma_i^{-2})^{-0.5}$.  For the strongest C~I lines, typical saturation
corrections arising from the curves of growth were as follows: 0.05~dex
for $\gamma$~Ori, 0.01~dex for $\lambda$~Lup, 0.27~dex (thermal) or
0.51~dex (turbulent) for $\delta$~Cyg, and 0.08~dex for $\alpha$~Del.

\placetable{col_dens}
\begin{deluxetable}{
r    
c    
c    
c    
c    
c    
}
\tablecolumns{6}
\tablewidth{0pt}
\tablecaption{Column Densities and $f1$ Values\label{col_dens}}
\tablehead{\colhead{Star} & \colhead{Line} & \colhead{Assumed $T$} &
\colhead{$N$(C~I)} & \colhead{$N$(C~I*)} &
\colhead{$f1$\tablenotemark{b}}\\
\colhead{Name} & \colhead{Broadening\tablenotemark{a}} & \colhead{(K)} &
\colhead{($10^{12}\,{\rm cm}^{-2}$)} &
\colhead{($10^{12}\,{\rm cm}^{-2}$)}
}
\startdata
$\gamma$~Ori&thermal&152&
$1.13\pm 0.12$&$0.104\pm 0.171$&$0.083^{+0.124}_{-0.155}$\\
&turbulent&$\ll 152$&
$1.17\pm 0.14$&$0.105\pm 0.171$&$0.081^{+0.124}_{-0.150}$\\
$\lambda$~Lup&thermal&5560&
$1.10\pm 0.16$&$0.251\pm 0.177$&$0.179^{+0.122}_{-0.125}$\\
&turbulent&$\ll 5560$&
$1.11\pm 0.17$&$0.251\pm 0.177$&$0.179^{+0.122}_{-0.125}$\\
$\delta$~Cyg&thermal&245&
$6.78^{+1.20}_{-1.02}$&$1.27\pm 0.40$&$0.154^{+0.064}_{-0.057}$\\
&turbulent&$\ll 245$&
$11.7^{+4.2}_{-2.9}$&$1.36\pm 0.45$&$0.103^{+0.063}_{-0.049}$\\
$\alpha$~Del&thermal&1680&
$1.91\pm 0.13$&$0.51\pm 0.18$&$0.204^{+0.064}_{-0.068}$\\
&turbulent&$\ll 1680$&
$2.07\pm 0.14$&$0.51\pm 0.18$&$0.192^{+0.061}_{-0.065}$\\
\enddata
\tablenotetext{a}{Calculations were performed for two possible extremes in the
assumed line broadening of the C~I features: either the lines broadened by thermal
Doppler motions for a temperature given by Eq.~\protect\ref{T_NaI} (see the
numerical values in the next column) or they are broadened purely by turbulence
with arbitrarily low gas temperatures.}
\tablenotetext{b}{$f1\equiv N$(C~I*)/$N$(C~I$_{\rm total}$).  Absorption features
from the second excited level, $N$(C~I**), are not observable because they are too
weak.  Hence, a value of $N$(C~I**)/$N$(C~I*)=0.15 is assumed to arrive at a value
for $N$(C~I$_{\rm total}$).  See text for details.}
\end{deluxetable}

\subsection{Fine-Structure Excitations}\label{fsl}

A conventional means of expressing the fine-structure excitation of C~I
is through the quantities $f1\equiv N({\rm C~I^*})/N({\rm C~I}_{\rm
total})$ and $f2\equiv N({\rm C~I^{**}})/N({\rm C~I}_{\rm total})$ 
(Jenkins \& Shaya 1979; Jenkins et al. 1981, 1998; Jenkins, Jura, \&
Loewenstein 1983; Smith et al. 1991; Jenkins \& Wallerstein 1995).  The
utility of this representation arises from the ease of interpreting
discrepancies in the excitations of the two levels in terms of a
superposition of contributions from regions with contrasting physical
conditions  (Jenkins \& Tripp 2001).  Since C~I* was the only excited
level that created strong enough lines to measure the present survey,
the special advantage of the $f1,f2$ representation is lost. 
Nevertheless, in the interest of maintaining consistency with the
earlier studies of conditions elsewhere, there is some merit in
retaining  the $f1$ representation in favor of the more straightforward
quantity $N$(C~I*)/$N$(C~I).

To evaluate $f1$, one needs to estimate the amount of C~I** present,
since it makes a small contribution to C~I$_{\rm total}$.  Calculations
of the expected level populations indicate that $N({\rm
C~I^{**}})\approx 0.15N({\rm C~I^*})$ when the excitations are low and
there is no complex mixture of gases with very different conditions
[e.g., see the theoretical tracks in Fig.~6 of Jenkins \& Tripp 
(2001)].  Thus, values of $f1$ listed in the last column of
Table~\ref{col_dens} were evaluated from the expression
\begin{equation}\label{f1}
f1={N({\rm C~I^*})\over N({\rm C~I})+1.15N({\rm C~I^*})}~.
\end{equation}
Upper or lower limits for $f1$ represent the outcomes from Eq.~\ref{f1}
using opposite 1-$\sigma$ extremes of the permitted column densities of
$N$(C~I) and $N$(C~I*).

\section{Interpretation}\label{interpretation}

\subsection{Expected Values of $f1$}\label{expected_f1}

The next step in the investigation is to explore the expected behavior
of $f1$ over a broad range of possible physical conditions.  The
equilibrium concentrations of the atoms in different fine-structure
excitations are governed by the balance between collisional excitations,
collisional de-excitations, and the levels' rates of spontaneous
radiative decay.  Small adjustments arise also from optical pumping of
the levels  (Jenkins \& Shaya 1979).   The calculations adopted here
duplicate those described by Jenkins \& Tripp  (2001), except for an
additional task of estimating the fractional ionization of the gas in
the regime of high temperatures and low densities.  Knowing the degree
of ionization is important because electrons and protons have collision
rate constants that are appreciably higher than those of neutral
hydrogen atoms when temperatures are much above 100$\,$K  (Keenan 1989).

The diffuse, neutral interstellar medium is partly ionized by cosmic
rays (at an estimated rate $\xi_{\rm CR}=3\times 10^{-17}\,{\rm s}^{-1}$
that includes ionizations from secondary particles) and EUV and soft
x-ray photons (at a rate $\xi_{\rm XR}$) that manage to penetrate
through an absorbing layer of neutral H and He near the edges of a
region  (Wolfire et al. 1995).  The fractional ionization of H is
governed by the balance between these ionizing processes and
recombinations,
\begin{equation}\label{H_ioniz_equilib}
n({\rm H~I})(\xi_{\rm CR}+\xi_{\rm XR})=\alpha_{2,{\rm
H}}n(e)^2+\alpha_{g,{\rm H}}n(H)n(e)
\end{equation}
where the right-hand side of the equation includes two recombination
channels for the protons: (1) recombinations with free electrons to all
energy states $n=2$ and higher and (2) recombinations with negatively
charged, small grains and polycyclic aromatic hydrocarbon molecules 
(Weingartner \& Draine 2001), with a rate constant $\alpha_{g,{\rm H}}$
normalized to the total hydrogen density $n(H)$.  If we assume the
shielding of the inner region of a cloud arises from an external column
density $N({\rm H~I})\sim 10^{19}\,{\rm cm}^{-2}$ (an approximately
correct value for the regions viewed here; see \S\ref{temp_limits}), the
calculations of Wolfire et al  (1995) indicate that $\xi_{\rm
XR}=7\times 10^{-17}\,{\rm s}^{-1}$ for $n(e)\sim 10^{-2}\,{\rm
cm}^{-3}$.  Supplementing the free electrons from the ionization of H
(and He) are those created from elements that are almost completely
ionized because they have ionization potentials less than 13.5$\,$eV. 
This contribution is about $2\times 10^{-4}n({\rm H})$.  

\begin{figure}
\plotone{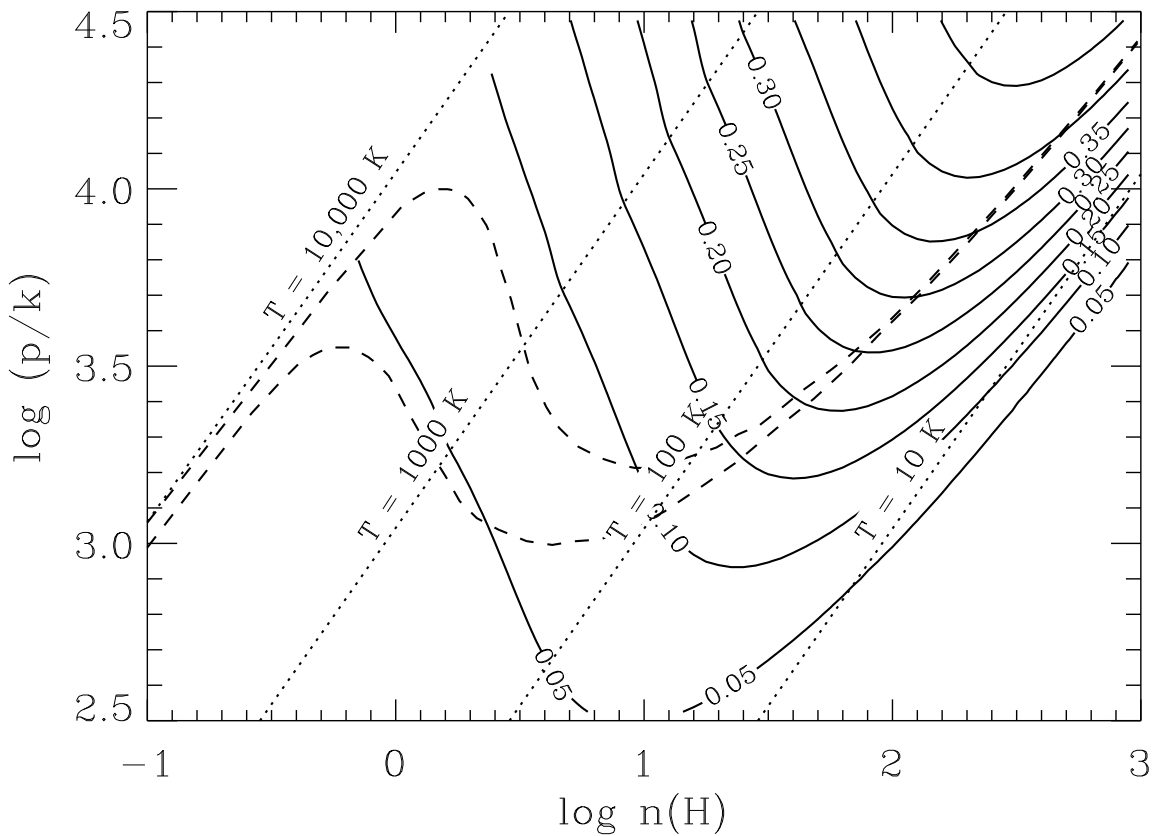}
\caption{Expected values of $f1\equiv N({\rm C~I^*})/N({\rm C~I}_{\rm total})$
[solid contour levels] as a function of the parameters $\log (p/k)$ and $\log
n({\rm H})$.  Diagonal straight, dotted lines show the loci of constant
temperatures $T$ with labels indicating the values.  Curved, dashed lines show the
thermal equilibrium curves derived by Wolfire et al.  (1995) for two different
amounts of shielding of EUV and soft x-ray radiation (upper curve: $N({\rm H~I})=
10^{18}\,{\rm cm}^{-2}$; lower curve:  $N({\rm H~I})= 10^{19}\,{\rm
cm}^{-2}$).\label{f1_conts}}
\end{figure}

Figure~\ref{f1_conts} shows the outcomes for the expected values of $f1$
in the form of a contour diagram in a representation of $\log (p/k)$ vs.
$\log n({\rm H})$.  It is clear that without independent information
that constrains either $T$ or $n({\rm H})$, measurements of $f1$ are of
little value in defining the thermal pressure $p$ (except for defining
lower bounds when there is a well defined lower limit for $f1$).  In the
following discussions, we consider different ways to constrain the
allowed temperatures in the C~I-bearing regions.  Even rudimentary
limits can be beneficial.

\placefigure{f1_conts}

\subsection{Temperature Ranges}\label{temp_limits}

\subsubsection{Upper Limits}\label{upper_T_limits}

The most straightforward upper limits for $T$ arise from the narrowness
of the Na~I lines seen in the spectra of $\gamma$~Ori and $\delta$~Cyg. 
The application of Eq.~\ref{T_NaI} to the $b$-values given in
\S\ref{col_densities} yields upper limits $T=152$ and 245$\,$K for the
respective stars.  These temperatures are located only slightly above
the sequence of positions where the $f1$ contours loop down to their
minima in $\log (p/k)$ in Fig.~\ref{f1_conts}.   Unfortunately, the
broader lines for Na~I toward $\lambda$~Lup and $\alpha$~Del could allow
much higher values of $T$, i.e., ones that do not sufficiently constrain
the allowed $p/k$ much below $10^{4.5}\,{\rm cm}^{-3}\,$K for the
measured values of $f1$ toward these stars.  Additional means for
constraining the allowed temperatures must rely on some independent
theoretical arguments.

The tactic for determining the largest probable temperatures of material
in front of $\lambda$~Lup and $\alpha$~Del will be to prove that above a
certain temperature, it is difficult to produce the observed amount of
C~I.  In order to do this, we must start with an estimate of the amount
of gas present in each of the C~I-bearing clouds.  It is generally
acknowledged that the abundance of sulfur atoms in the interstellar
medium is close to the solar abundance ratio relative to hydrogen, ${\rm
S/H} = 1.9\times 10^{-5}$  (Federman et al. 1993; Spitzer \& Fitzpatrick
1993; Fitzpatrick \& Spitzer 1994, 1997; Howk, Savage, \& Fabian 1999),
i.e., sulfur is not appreciably depleted onto dust grains.  Sulfur in
its singly ionized form is the most abundant stage expected for H~I
regions.  Since S~II can also reside in H~II regions (its ionization
potential equals 23.4~eV) a determination of $N$(H~I) using $N$(S~II)
is, strictly speaking, only an upper limit.  However the arguments about
not producing enough ${\rm C~I_{\rm total}}$ to match the observations
will only be strengthened if $N$(H~I) is really less than this limit. 
Thus we can simplify the discussions and adopt a conservative position
by treating the estimate for $N$(H~I) as an actual value.

\begin{figure}
\plotone{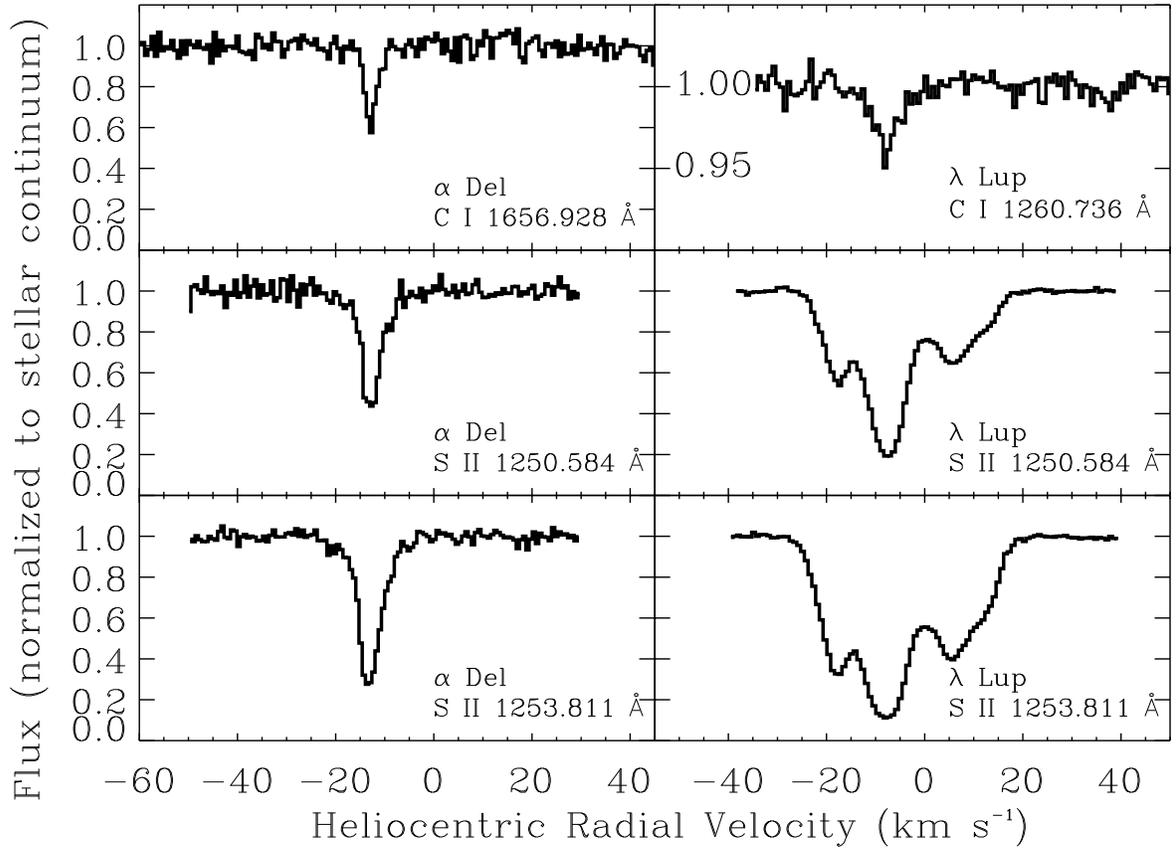}
\caption{Absorption profiles for various transitions of C~I and S~II, as
indicated, in the spectra of $\alpha$~Del (left) and $\lambda$~Lup (right).  The
intensities have been normalized to the stellar continua. Note that the scale for
the C~I transition for $\lambda$~Lup has been expanded.\label{sii_plts}}
\end{figure}

The middle and lowest panels of Figure~\ref{sii_plts} show the two
weakest absorption features of S~II in the spectra of $\alpha$~Del and
$\lambda$~Lup.  The ratios of the $f$-values of the two transitions is
2.0  (Morton 1991), but it is clear from the plots that the ratios of
the apparent optical depths in the stronger parts of the profiles are
smaller than 2.  While one way to determine $N$(S~II) is to measure the
equivalent widths of the lines and use the standard curve of growth
analysis, a better approach is to measure the apparent optical depth
ratios of the two lines at every velocity and then apply a method
described by Jenkins  (1996) to correct for distortions in the weak line
caused by unresolved, saturated structures within the profile.

\placefigure{sii_plts}

For $\alpha$~Del, $N({\rm S~II})=3.1\times 10^{14}\,{\rm cm}^{-2}$.  In
turn, one infers that $N({\rm H~I})=1.6\times 10^{19}\,{\rm cm}^{-2}$
and $N({\rm C~II})=2.2\times 10^{15}\,{\rm cm}^{-2}$ if the abundance of
carbon relative to hydrogen is the typical value of about $1.4\times
10^{-4}$ in the interstellar medium  (Hobbs, York, \& Oegerle 1982;
Cardelli et al. 1991, 1993, 1996; Sofia et al. 1997; Sofia, Fitzpatrick,
\& Meyer 1998).  Since the ionization and recombination properties of
S~II and C~II are very similar  (Sofia \& Jenkins 1998; Howk \& Sembach
1999), shifts in the ratio of these two first ions caused by ionization
in low density, partly ionized gas should not be a problem.  Comparing
the neutral carbon column densities toward $\alpha$~Del reported in
Table~\ref{col_dens} to the inferred $N$(C~II) derived here, we arrive
at $\log N({\rm C~I}_{\rm total})-\log N({\rm C~II})=-2.97$.  The
significance of this number will be evident when it is compared to
theoretical expectations under many different conditions.

\begin{figure}
\plotone{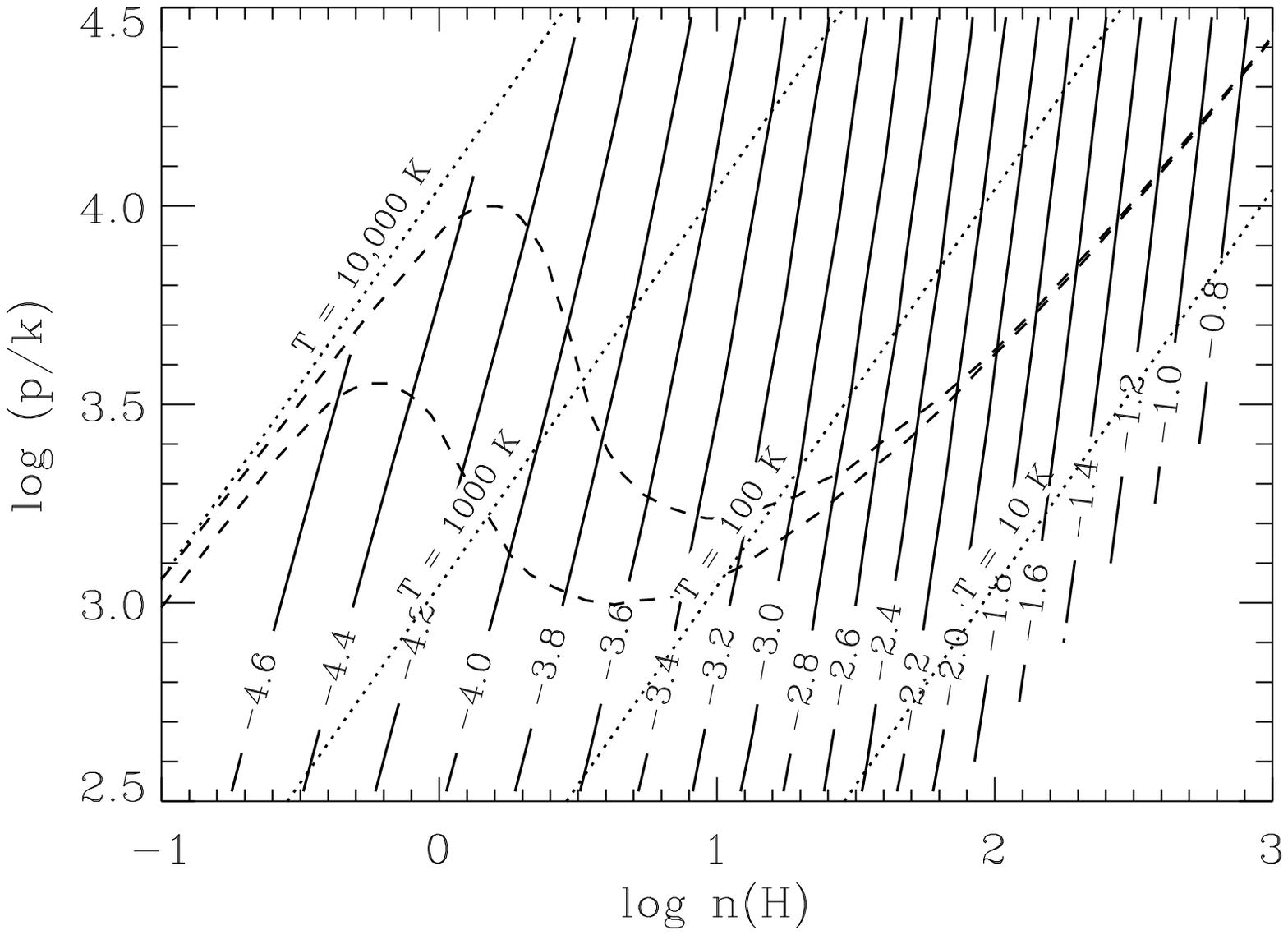}
\caption{Expected values of $\log n({\rm C~I_{\rm total}})-\log n({\rm C~II})$ as
a function of the parameters $\log (p/k)$ and $\log n({\rm H})$, found by solving
Eq.~\protect\ref{C_ioniz_equilib}.  Other lines in the diagram are described in
the caption for Fig.~\protect\ref{f1_conts}.\label{c_conts}}
\end{figure}

Figure~\ref{c_conts} shows the contours for the expected values of $\log
n({\rm C~I_{\rm total}})-\log n({\rm C~II})$ arising from the
equilibrium condition
\begin{equation}\label{C_ioniz_equilib}
n({\rm C~I_{\rm total}})\xi_{\rm C~I}=n({\rm C~II})[n(e)\alpha_{\rm
C}+\alpha_{g,{\rm C}}n({\rm H})]
\end{equation}
where the assumed interstellar ionization rate $\xi_{\rm C~I}=2.24\times
10^{-10}\,{\rm s}^{-1}$  (Jenkins \& Shaya 1979), the recombination rate
of singly-ionized carbon with free electrons $\alpha_{\rm C}$ is
evaluated from the fitting equation given by  Shull \& van Steenberg 
(1982), and the grain/PAH recombination rate $\alpha_{g,{\rm C}}$ is
from the relation specified by Weingartner \& Draine  (2001) with their
radiation field strength parameter $G$ set to 1.13.  Electron densities
$n(e)$ arise from solutions to Eq.~\ref{H_ioniz_equilib}, as described
earlier.  [If one neglects grain recombination altogether, the predicted
values for $n({\rm C~I_{\rm total}})/n({\rm C~II})$ are not much
different because the reduced total recombination rate is compensated by
a larger predicted value for $n(e)$.]

\placefigure{c_conts}

By comparing the contours in Figs.~\ref{f1_conts} and \ref{c_conts}, it
is clear that if $\alpha$~Del were to have $f1$ equal to the nominal
value of 0.20 (for pure thermal broadening of the lines), the
temperature must be equal to or less than 150$\,$K to give an expected
yield of ${\rm C~I_{\rm total}}$ that is at least as large as the
observed value.  If $f1$ is as high as the $+1\sigma$ limit, the
temperature could be as high as 250$\,$K.  The upper limit for $T$ if
$f1$ is at its $-1\sigma$ limit is 90$\,$K.  These temperature limits
are substantially lower than those set by the width of the Na~I
absorption feature.

For $\lambda$~Lup, the strong, central peak of the S~II absorption
matches the velocity of the C~I profile (see Fig.~\ref{sii_plts}).  The
S~II associated with this peak has a column density $N({\rm
S~II})=1.2\times 10^{15}\,{\rm cm}^{-2}$. (It is reasonable to consider
the gas giving absorptions at velocities on either side of this peak to
arise from unrelated material.)  Using reasoning identical to that
applied to $\alpha$~Del, one finds for $\lambda$~Lup that $N({\rm
H~I})=6.2\times 10^{19}\,{\rm cm}^{-2}$, $N({\rm C~II})=8.7\times
10^{15}\,{\rm cm}^{-2}$, and  $\log N({\rm C~I}_{\rm total})-\log N({\rm
C~II})=-3.81$.  Clearly, the restriction on $T$ for $\lambda$~Lup is not
as stringent as for $\alpha$~Del: a limit $T<2700\,$K arises from the
interception of a line corresponding to $f1=0.19$ (nominal value) with
the expectation $\log N({\rm C~I}_{\rm total})-\log N({\rm
C~II})=-3.81$.  This limit is only mildly more restrictive than the one
that arises from the line width of Na~I.  For $f1=0.05$ (the $-1\sigma$
limit) the limiting temperature drops to 350$\,$K.

\subsubsection{Lower Limits}\label{lower_T_limits}

A task which now remains is to confine the allowable temperatures on the
low side, since, as with high temperatures, arbitrarily high pressures
could occur if $T$ were permitted to be arbitrarily low (again, see
Fig.~\ref{f1_conts}).  In order to realize such a constraint, one may
consider the sequence of temperatures at different densities predicted
for thermal equilibrium in the interstellar medium.  Wolfire et al. 
(1995) calculated equilibria for the diffuse phases of interstellar
material for various assumptions.  The two curved, dashed lines in
Fig.~\ref{f1_conts} show the outcomes for their predictions based on
heating rates for material shielded by absorbing columns $N({\rm H~I})=
10^{18}$ (upper curve) and $10^{19}\,{\rm cm}^{-2}$ (lower curve), using
cooling rates based on normal grain and heavy element abundances. 
According to Field's  (1965) thermal stability criterion, stable phases
of the interstellar medium should be located only along portions of the
equilibrium curves that have a positive slope in the $\log p$ vs. $\log
n$ diagram  (Field, Goldsmith, \& Habing 1969; Shull 1987; Begelman
1990).  In a simplistic application of this principle, one would predict
that stable phases would exist only at slightly below 10,000$\,$K
(left-hand branch) or in the range $30<T<300\,$K (right-hand branch). 
In reality, this picture is only partly correct.  On the one hand,
observations of 21-cm emission and absorption in directions toward
extragalactic sources indicate that about half of the H~I in the diffuse
medium violates this condition  (Heiles 2001).  The data show strong
evidence that gas can often be found at temperatures that are
intermediate between the two (positive-slope) equilibrium curves, an
effect that is probably attributable to the effects of mixing of the two
extremes in a regime strongly dominated by turbulence  (V\'azquez-Semadeni, Gazol, \& Scalo 2000; Gazol et al. 2001).  On the other hand,
the observations generally show that the lowest H~I temperatures
coincide with the theoretical expectations.  While broad surveys of
emission with superposed absorption highlight a few, selected locations
where spin temperatures are below 30$\,$K   (Gibson et al. 2000; Knee \&
Brunt 2001), the sight lines toward extragalactic sources offer a chance
to sample random volumes: the survey of emission and absorption toward
these sources reported by Heiles  (2001) and Heiles \& Troland  (2002)
reveals that only about 3.7\% of the cold, neutral medium is at
temperatures below $25\,$K.  In short, the 21-cm data indicate that a
good working assumption is that $T$ is usually above the right-hand
branch of the thermal equilibrium curve, and violations of the lower
limit defined by this curve are rare.  For this reason, it is reasonable
to regard the nearly coincident dashed lines on the right-hand side of
Fig.~\ref{f1_conts} as a good representation for the lowest probable
temperatures for different pressures.

\section{Conclusions and Possible Interpretations}\label{summary}

\begin{figure}
\plotone{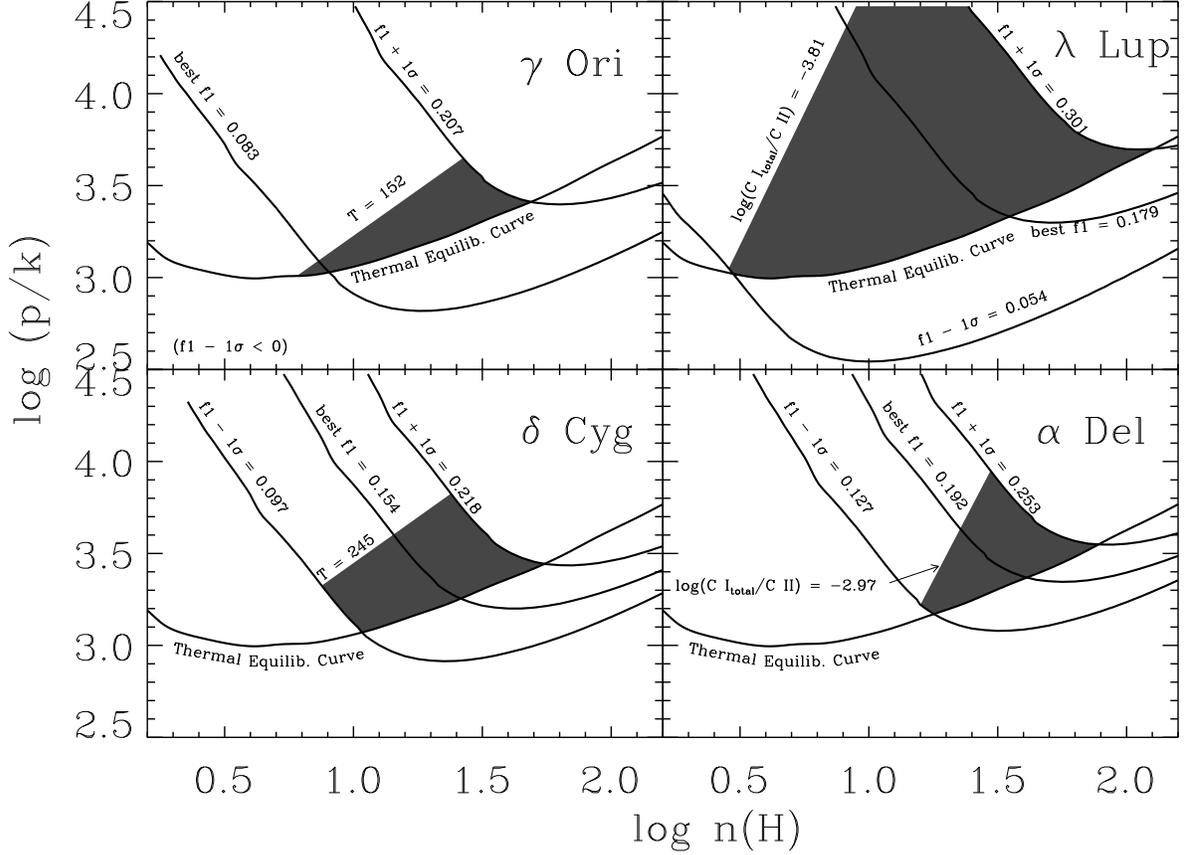}
\caption{Allowed combinations of $\log (p/k)$ and $\log[n({\rm H})]$ permitted by
the $\pm 1\sigma$ error limits for $f1$ for the C~I-absorbing regions in front of
the four targets.  Intersecting constraints arise from the thermal equilibrium
curve for an external absorbing column $N({\rm H~I})=10^{19}\,{\rm cm}^{-2}$
calculated by Wolfire, et al  (1995), and temperature limits defined either from
the Doppler width of Na~I absorption lines or, alternatively, the requirement that
the expected yield of ${\rm C~I_{total}}$ relative to C~II from the condition of
ionization equilibrium (Eq.~\protect\ref{C_ioniz_equilib}) is as least as much as
actually observed.  General layouts of the expected C~I fine-structure population
ratios and ${\rm C~I_{\rm total}/C~II}$ ionization equilibria are shown in
Figs.~\protect\ref{f1_conts} and \protect\ref{c_conts},
respectively.\label{limit_panels}}
\end{figure}

A synthesis of the conclusions presented earlier is shown in
Figure~\ref{limit_panels}, again in the representation of $\log (p/k)$
vs. $\log n({\rm H})$.  To constrain the allowed physical conditions,
the outcomes presented in Table~\ref{col_dens} for the fine-structure
population ratios of C~I, represented by the parameter $f1$, must be
supplemented by intersecting lines representing temperature limits
derived in \S\ref{temp_limits}.  Of these limits, the upper bounds
derived from the measurements of the Na~I line widths by other
investigators are the most direct and reliable.  However these
measurements are valuable only for the cases of $\gamma$~Ori and
$\delta$~Cyg.  Alternative means for limiting $T$, such as using the
thermal equilibrium curve to define a lower bound and the carbon
ionization equilibrium for an upper bound, are indirect and subject to
less certain theoretical assumptions.

\placefigure{limit_panels}

It should be noted that for each case the lines that border the allowed
(shaded) regions do not enclose the worst possible deviations.  For
instance, the $f1$ limits represent only $\pm 1\sigma$ deviations.  The
temperature constraints from the width of the Na~I absorption lines do
not include possible errors in measurement, since they are difficult to
assess.  The theoretical limits based on thermal equilibria or the
expected fractional abundances of carbon in neutral form are only as
good as the assumptions that were incorporated into their respective
developments.  In spite of these shortcomings, when three of the stars
(those other than $\lambda$~Lup) are examined collectively, they
indicate that clouds within the Local Bubble have thermal pressures in
the range $10^3<p/k<10^4\,{\rm cm}^{-3}\,$K.

From the introductory remarks presented in \S\ref{intro}, it is clear
that the most interesting limit on pressure is the upper one.  Any
single case is not compelling enough to support the proposal that
$p/k<10^4\,{\rm cm}^{-3}\,$K, since deviations in excess of $1\sigma$ in
a particular direction can be expected to occur 16\% of the time. 
Nevertheless, if one supposes that all clouds within the Local Bubble
have a single representative pressure, it is difficult to imagine that
all of the random observational errors would conspire to mislead us to
believe that $p/k<10^4\,{\rm cm}^{-3}\,$K when in reality the pressure
is much higher than this value.  The result for $\lambda$~Lup is not
helpful in constraining $p$ on the high side, but at least it shows an
outcome on the low side that is consistent with those of the other
stars.  As for systematic errors, the only easily identifiable ones are
possible errors in the relative $f$-values of the C~I and C~I*
transitions or errors in the parameters used to develop the predicted
values of $f1$, such as the collisional rate constants or spontaneous
radiative decay rates.

It is apparent from this study that a seemingly large disparity in
pressure between neutral clouds embedded in the Local Bubble and the
surrounding hot gas is a general phenomenon and not just confined to the
LIC around the Sun.  Possible solutions to the dilemma posed by the
apparent pressure imbalance may arise from the following considerations:
\begin{enumerate}
\item Perhaps the clouds have not had enough time to respond fully to a
sudden increase in external pressure that happened during the event(s)
that created the LB, perhaps some $10^5-10^7$~yr ago  (Cox \& Reynolds
1987).  If this is the case, each cloud is in an intermediate condition
where a shock is still progressing from the periphery of the cloud,
perhaps well outside the portion containing the C~I, to its center. 
However, this explanation is not satisfactory for the LIC, since the
distance from the Sun to the closest boundary divided by the sound speed
is less than $10^4$~yr.
\item Perhaps magnetic fields within the clouds provide the support
needed to counteract the high pressure from the outside.  An upper limit
$B<3\mu$G for the magnetic field just outside the heliosphere 
 (Gloeckler, Fisk, \& Geiss 1997) diminishes the attractiveness of this
explanation in the case of the LIC.
\item Additional contributions to the clouds' internal pressures might
arise from turbulent motions.  However, the narrowness of the Na~I lines
observed for $\gamma$~Ori and $\delta$~Cyg lowers the prospects for this
proposal.  For instance, if one uses the limit $n({\rm
H})<10^{1.7}\,{\rm cm}^{-3}$ shown in Fig.~\ref{limit_panels} and
assumes the Na~I lines are broadened only by turbulence, one finds that
$p_{\rm turb}/k=\rho b^2/2k < 470$ and $800\,{\rm cm}^{-3}\,$K for
$\gamma$~Ori and $\delta$~Cyg, respectively.
\item Perhaps the measurements of the thermal pressure for the hot gas
in the Local Bubble were incorrect because they relied on either
inaccurate models for the line emission or the assumption that the
radiation from the hot plasma is emitted under equilibrium conditions. 
For a plasma that is cooling very rapidly, perhaps through adiabatic
expansion, the x-rays may come principally from delayed recombinations
in cooler gas, rather than collisional excitations at higher
temperatures  (Breitschwerdt \& Schmutzler 1994).  Apart from this
theoretical proposal, there have also been persistent difficulties in
reconciling measurements of x-ray and EUV line emission with models for
a hot plasma in collisional equilibrium and with cosmic abundance ratios
of the heavier elements  (Jelinsky, Vallerga, \& Edelstein 1995;
McCammon et al. 2002).  In trying to understand the soft x-ray spectrum
measured by the Diffuse X-ray Spectrometer (DXS), Sanders et al.  (2001)
even explored models with depleted element abundances and nonequilibrium
conditions arising from either rapid heating or cooling.  From their
inability to achieve reasonable fits of these more complex models with
their data, they suggested that the discrepancies might arise from
inaccurate atomic data which were used to predict the expected emission
line strengths.  If these models are in need of revision, then one might
question the earlier estimates of hot gas emission measures from low
resolution surveys of x-ray emission that led to the finding that $p/k >
10^4\,{\rm cm}^{-3}$K within the LB.
\end{enumerate}

\acknowledgements

The author thanks J.~Weingartner for supplying a program that calculated
the partial ionization of the ISM, as described in \S\ref{expected_f1}. 
Important parameters for the target stars were obtained from the VizieR
web site  (Ochsenbein, Bauer, \& Marcout 2000), as noted in the text and
Table~\ref{target_stars}.  Support for this research was provided by
NASA through grant number GO-6415.01-95A from the Space Telescope
Science Institute, which is operated by the Association of Universities
for Research in Astronomy, Inc., under NASA contract NAS5-26555.

\appendix
\section{Fixed-Pattern Noise Analysis}\label{fpn_analysis}
\subsection{Equations}\label{equations}

The discussion here addresses the problem of disentangling the
detector's fixed pattern response function\footnote{As noted in
\S\protect\ref{artifacts}, for GHRS the fine-scale sensitivity pattern
is not always stationary with respect to the readout diode array.  One
must explicitly determine the small shifts in position of the
photocathode's pattern function for each exposure, since the mapping of
the photocathode onto the readout diodes can change with time.  In the
discussion presented here, the term ``stationary'' means stationary with
respect to the photocathode's reference frame, which may or may not move
across the diodes from one exposure to the next, depending on the
observing circumstances.  The notion of a movement refers to the
intentional displacement of the spectrum caused by a commanded motion of
the grating carousel.} from the spectrum that moves, on the basis of
information presented in the separate exposures. The analysis method
solves for the two functions, but with the actual work being carried out
in the Fourier Transform domain.  This method has two important
features.  First, it offers a direct evaluation of the solutions, rather
than relying on an initial guess for the two functions that is followed
by iterations that gradually improve the representations until they
converge toward consistent representations.  Second, by exploring the
properties of equations that operate in the Fourier domain, we gain
insights on some near singular conditions that, in some practical
circumstances, give poorly defined results when the shift dimensions
have common factors in wavelength.  This problem, discussed in
\S\ref{limitations}, warns us of some potential dangers of programming
shifts that are of uniform magnitude, as was intended for the FP-SPLIT
routines on GHRS.  

An individual spectrum $F(\lambda)$ recorded by a spectrograph consists
of the actual spectrum $A(\lambda)$ multiplied by the pattern function
$P(\lambda)$, plus a noise contribution arising from statistical
fluctuations in the recorded photoevents.  For the Digicon detector on
GHRS, $P(\lambda)$ has the appearance of a function that, over large
scales, is nearly constant, but with a high frequency granularity having
an amplitude of a few percent.  However, occasionally there are flaws in
the photocathode that give much larger spikes  (Ebbets 1992). In the
reference frame of the detector's photocathode, separate recordings of
$F_i(\lambda)$ using the FP-SPLIT routine consist of the stationary
pattern function $P(\lambda)$ with different superpositions of the
actual spectrum $A(\lambda)$ having offsets with magnitudes $\delta_i$
in the wavelength direction, where the subscript $i$ differentiates the
separate exposures.  The magnitudes of $\delta_i$ may be determined by
either cross correlating the exposures (for a spectrum with many narrow
features) or comparing the offsets of positions of a strong line.  The
latter is recommended over the former if a single, narrow feature is
present, with the rest of the spectrum being mostly random noise or much
broader features.

As a first step in the analysis, we can reformulate the observed
spectrum $F(\lambda)$ in terms of its logarithm, so that the mixing of
two functions represented by $P(\lambda)$ and $A(\lambda)$ is a simple
linear sum, $\log[P(\lambda)] + \log[A(\lambda - \delta_i)]$.  A
restriction that should be imposed is that downward excursions of both
$P(\lambda)$ and $A(\lambda)$ are not very large compared with the
average signal levels.  If this condition is violated, the negative dips
can have a disproportionately large influence on the outcome, especially
because the analysis works in the Fourier domain where large (negative)
$\delta$-functions can have a global effect.  For $P(\lambda)$ this
restriction to small amplitudes is generally satisfied for GHRS, except
at the locations of serious flaws in the photodiode array (which are
normally corrected out beforehand).  For the actual spectrum
$A(\lambda)$, we require that there are no very deep absorption
features.  (Methods for overcoming problems with deep features in the
spectrum will be discussed very briefly in \S\ref{limitations}.)

In the Fourier domain we define the real and imaginary parts of the
pattern's transform by the functions in inverse wavelength $\omega$, 
\begin{equation}\label{p_x}
p_x(\omega) = \Re \bigl\{ {\cal F}[\log P(\lambda)]\bigr\}
\end{equation}
and
\begin{equation}\label{p_y}
p_y(\omega) = \Im \bigl\{ {\cal F}[\log P(\lambda)]\bigr\}\ ,
\end{equation}
where $\cal F$ is the Fourier transform operator.  For $F(\lambda)$ and
$A(\lambda)$ we designate the functions $f_x$, $f_y$, $a_x$ and $a_y$ in
the same manner and allow for the fact that separate recordings will
have different phase shifts for $a$ that are invoked by multiplying the
zero-shift, complex function $a_x(\omega) + ia_y(\omega)$ by a factor
$\exp(-2\pi i\omega\delta_i)$.

The best solutions for $p_x(\omega)$, $p_y(\omega)$, $a_x(\omega)$, and
$a_y(\omega)$ may be evaluated at each $\omega$ independently by 
solving for the smallest possible sum $Q^2$ of the squared real and
imaginary residuals
\begin{equation}\label{q2x}
Q^2_x = \sum_{i=1}^n(a_x\cdi+a_y\sdi+p_x-f_{x,i})^2
\end{equation}
and
\begin{equation}\label{q2y}
Q^2_y = \sum_{i=1}^n(-a_x\sdi+a_y\cdi+p_y-f_{y,i})^2~,
\end{equation}
where $\Delta_i$ is an abbreviation for $2\pi\omega\delta_i$.  This
minimum is achieved by setting the partial differentials of $Q^2$ with
respect to the four unknowns $a_x$, $a_y$, $p_x$, and $p_y$ equal to
zero.  If ${\bf u}(\omega)$ represents a vector of these respective
unknowns at each $\omega$, we must solve a system of 4 linear equations 
\begin{equation}\label{linear_eq}
{\bf C}(\omega)\cdot{\bf u}(\omega)={\bf b}(\omega)
\end{equation}
where the coefficients of $C_{i,j}$ of the matrix {\bf C} and $b_i$ of
the vector {\bf b} are listed in Table~\ref{linear_eq_coeffs}.

\placetable{linear_eq_coeffs}
\begin{deluxetable}{
c    
c    
c    
c    
c    
c    
c    
}
\tablecolumns{7}
\tablewidth{0pt}
\tablecaption{Coefficients\tablenotemark{a}\ \ for the Matrix and Vector in
Equation~\ref{linear_eq}\label{linear_eq_coeffs}}
\tablehead{\colhead{Term} & \multicolumn{5}{c}{$C_{i,j}$} & \colhead{$b_i$}\\
\colhead{} & \colhead{$i$} & \colhead{$j=1$} & \colhead{$j=2$} & \colhead{$j=3$} &
\colhead{$j=4$} & \colhead{}}
\startdata
\partials{a}{x}&1&$n$&0&$\sum\cdi$&$-\sum\sdi$&$\sum f_{x,i}\cdi-\sum
f_{y,i}\sdi$\\
\partials{a}{y}&2&0&$n$&$\sum\sdi$&$\sum\cdi$&$\sum f_{x,i}\sdi+\sum
f_{y,i}\cdi$\\
\partials{p}{x}&3&$\sum\cdi$&$\sum\sdi$&$n$&0&$\sum f_{x,i}$\\
\partials{p}{y}&4&$-\sum\sdi$&$\sum\cdi$&0&$n$&$\sum f_{y,i}$\\
\enddata
\tablenotetext{a}{Each sum is evaluated from $i=1$ to $n$, where $n$ is the number
of independent spectral recordings.}  
\end{deluxetable}

After solving for each vector ${\bf u}(\omega)$ over all $\omega$, one
simply evaluates the inverse Fourier transforms of the two pairs of
terms, $u_1(\omega)+iu_2(\omega)$ and $u_3(\omega)+iu_4(\omega)$ to
recover the best representations of $A(\lambda)$ and $P(\lambda)$,
respectively.

\subsection{Limitations}\label{limitations}

Large increases in the magnitudes of terms in the error matrices $[{\bf
C}(\omega)]^{-1}$ at certain frequencies indicate places where the
solutions are ill-defined.  As one would expect, the outcomes for
frequencies whose inverses are much larger than the total span of the
$\delta_i$ are not well determined.  The solutions for $A(\lambda)$ and
$P(\lambda)$ can exhibit very broad, spurious undulations (of opposite
sign), as an outcome of the analysis program's attempt to reconcile
small differences in the low-frequency components in the observed
spectra arising simply from noise and subtle systematic errors.  As a
practical matter, it is reasonable to assign artificially a condition
that any low-frequency components of the observed spectra must arise
from either the source spectrum $A(\lambda)$ or the detector response
$P(\lambda)$.  A similar quandary arises for certain frequencies where
integral numbers of sinusoidal variations in the spectrum can exactly
match the spacings between all of the $\delta_i$.  One consequence of
this phenomenon is that the programmed intent of having equal step sizes
the FP-SPLIT routine for GHRS was a misguided one.\footnote{This
shortcoming was noted by the STIS Instrument Definition Team.  As a
result, the separation of the multiple slits for the FP-SPLIT option in
STIS are incommensurate with each other  (Leitherer 2001).} 
Fortunately, in practice the step sizes usually turn out to be not quite
equal, and this removes the degeneracies of the solutions.

The presence of deep absorption features in a spectrum poses special
difficulties.  At the locations of these features, the logarithms of
$F(\lambda)$ surge to very large negative values, becoming even
uncomputable in the bottoms of lines where the absorption is total.  One
must acknowledge that there is no way to determine $P(\lambda)$ at
locations where less than two of the spectra have non-zero fluxes. 
Inelegant but nevertheless workable methods of avoiding irregularities
in the computations include adding a small offset to the spectrum so
that zero intensities are never encountered or, alternatively,
eradicating the line and replacing it with a continuum intensity level.  

\end{document}